\documentclass[sigconf]{aamas}

\AtBeginDocument{%
  \providecommand\BibTeX{{%
    \normalfont B\kern-0.5em{\scshape i\kern-0.25em b}\kern-0.8em\TeX}}}

\usepackage[utf8]{inputenc}
\usepackage{hyperref}
\usepackage[acronym]{glossaries}
\usepackage{graphicx}
\usepackage{appendix}
\usepackage{algorithm}
\usepackage[noend]{algpseudocode}
\usepackage{tabularx}
\usepackage{booktabs}
\usepackage{soul}
\usepackage{framed}
\usepackage{wrapfig}
\usepackage{enumitem}
\usepackage{makecell}
\usepackage{collcell,array}
\usepackage{flushend} 

\setcopyright{ifaamas}  
\copyrightyear{2020} 
\acmYear{2020} 
\acmDOI{} 
\acmPrice{} 
\acmISBN{} 
\acmConference[AAMAS'20]{Proc.\@ of the 19th International Conference on Autonomous Agents and Multiagent Systems (AAMAS 2020)}{May 9--13, 2020}{Auckland, New Zealand}{B.~An, N.~Yorke-Smith, A.~El~Fallah~Seghrouchni, G.~Sukthankar (eds.)}  

\graphicspath{ {images/} }

\newglossaryentry{hydra}{name=HYDRA,description={a method for automatically designing algorithms to complement a portfolio~\cite{Xu2010}}}
\newglossaryentry{autofolio}{name=AutoFolio,description={an automatically configured algorithm selector~\cite{Lindauer2015AutoFolioSelector}}}

\newacronym{anac}{ANAC}{Automated Negotiating Agents Competition~\cite{Baarslag2012The2010}}
\newacronym{genius}{GENIUS}{General Environment for Negotiation with Intelligent multi-purpose Usage Simulation~\cite{Lin12}}
\newacronym{smac}{SMAC}{Sequential Model-based optimization for general Algorithm Configuration~\cite{Hutter2011SequentialOptimization}}
\newacronym{sat}{SAT}{Boolean Satisfiability}
\newacronym{asp}{ASP}{Answer Set Programming}
\newacronym{csp}{CSP}{Constraint Satisfaction Problem}
\newacronym{mip}{MIP}{Mixed Integer Programming}
\newacronym{aop}{AOP}{Stacked Alternating Offers Protocol~\cite{rubinstein1982perfect, osborne1994course}}
\newacronym{saop}{SAOP}{Stacked Alternating Offers Protocol~\cite{aydougan2017alternating}}
\newacronym{ps}{PS}{Partner Selection}
\newacronym{sfm}{SFM}{Smith Frequency Model~\cite{VanGalenLast2012}}
\newacronym{fyu}{FYU}{Full Yield Utility}
\newacronym{cr}{CR}{Concession Rate}
\newacronym{ar}{AR}{Average Rate}
\newacronym{dcp}{DCP}{Default Configuration Performance}
\newacronym{smbo}{SMBO}{Sequential Model-Based Optimisation~\cite{Hutter2011SequentialOptimization}}
\newacronym{batna}{BATNA}{Best Alternative To a Negotiation Agreement~\cite{Fisher1981GettingIn}}
\newacronym{slurm}{SLURM}{Simple Linux Utility for Resource Management~\cite{Yoo2003SLURM:Management}}
\newacronym{cov}{CoV}{Coefficient of Variance}
\DeclareMathOperator*{\argmax}{arg\,max}

\newenvironment{conditions}
    {\par where:\par\tabularx{\columnwidth}{>{\(}l<{\)} @{\({}: {}\)} >{\raggedright\arraybackslash}X}}
    {\endtabularx\par}

\newcommand{\AlgoParams}[2]{
    \foreach\x/\y [count=\i] in #1 {
        \Statex\ifnum\i=1 \textbf{#2} \fi \tabto*{2cm} \x \tabto*{3cm} \y
    }
}

{}
{}

\glsdisablehyper

\newcommand{\shiftdown}[1]{\smash{\raisebox{-.5\normalbaselineskip}{#1}}}
\newcolumntype{L}{>{\collectcell\shiftdown}l<{\endcollectcell}}

\begin{document}

\glsunsetall
\title{Automated Configuration of Negotiation Strategies}  



\author{Bram M. Renting}
\affiliation{%
 \institution{Delft University of Technology}
 \city{Delft} 
 \country{The Netherlands}}
\email{bramrenting@gmail.com}

\author{Holger H. Hoos}
\authornote{Alphabethical order due to equal contribution}
\affiliation{%
 \institution{Leiden University}
 \city{Leiden} 
 \country{The Netherlands}}
\email{H.H.Hoos@liacs.leidenuniv.nl}
\orcid{0000-0003-0629-0099}

\author{Catholijn M. Jonker}
\affiliation{%
 \institution{Delft University of Technology \& Leiden University}
 \city{Delft} 
 \country{The Netherlands}}
\authornotemark[1]
\email{C.M.Jonker@tudelft.nl}
\orcid{0000-0003-4780-7461}

\begin{abstract}
Bidding and acceptance strategies have a substantial impact on the outcome of negotiations in scenarios with linear additive and nonlinear utility functions.
Over the years, it has become clear that there is no single best strategy for all negotiation settings, yet many fixed strategies are still being developed. 
We envision a shift in the strategy design question from: What \textit{is} a good strategy?, towards: What \textit{could} be a good strategy?
For this purpose, we developed a method leveraging automated algorithm configuration to find the best strategies for a specific set of negotiation settings.
By empowering automated negotiating agents using automated algorithm configuration, we obtain a flexible negotiation agent that can be configured automatically for a rich space of opponents and negotiation scenarios.

To critically assess our approach, the agent was tested in an \gls{anac}-like bilateral automated negotiation tournament setting against past competitors.
We show that our automatically configured agent outperforms all other agents, with a 5.1\% increase in negotiation payoff compared to the next-best agent.
We note that without our agent in the tournament, the top-ranked agent wins by a margin of only 0.01\%.

\end{abstract}

\keywords{Automated Negotiation; Automated Algorithm Configuration; Negotiation Strategy} 

\maketitle
\glsresetall

\section{Introduction}
As of the 1980s, researchers have tried to design  algorithms (or software agents) that can assist or act on behalf of humans in negotiations. Early adopters in this field are Smith, Sycara, Robinson, Rosenschein and Klein~\cite{Smith1980,Sycara1988,Sycara-Cyranski1985,Robinson1990,Rosenschein:1986:RIC:15215,Klein1989}. 

In 2010, the \gls{genius} platform was created to provide a test-bed for evaluating new developments in the field of automated negotiation. Alongside, the \gls{anac} competition series was organized to stimulate the  development of negotiation algorithms in academia. 
Every year, \gls{anac} poses a new challenge for contestants to cope with.
Today, the combined effort of \gls{genius} and \gls{anac} has resulted in a standardized test-bed with more than 100 negotiating agents and negotiation scenarios that are readily accessible for research on automated negotiation~\cite{BaarslagANAC2010-2015}.

The negotiators are generally hard-coded software agents, based on a strategy with fixed parameters that are tuned at design time to optimize its behavior. The difficulty lies not in developing a negotiator, but in winning the competition, as both the configuration space and the space of negotiation scenarios are large, and the competing agents change every year.
 
This makes manual configuration on larger sets of negotiation instances tedious, time-consuming and impractical. 
Furthermore, note that evaluating a single strategy on a large set of negotiation scenarios takes too much time to be practical.

To avoid these difficulties, agents have been configured on smaller sets~\cite{Matos1998}. Attempts were made to automate this process, for example using genetic programming~\cite{Holland1992AdaptationIntelligence}, but again only on specific and simplified test sets. For instance, agents were only tested in one or two scenarios, or merely optimized  against themselves~\cite{Eymann2001,Dworman1996}. The resulting agents are highly specialized with unpredictable performance when negotiating outside of their comfort zone. No attempts have been reported at automating this configuration task on  large-scale, broad sets of negotiation scenarios and opponent strategies.

In this work, we present a solution for the automated algorithm configuration problem for automated negotiation on large problem sets. 
We recreate a negotiation agent from literature~\cite{lau2006evolutionary} that is configured manually, combine it with contemporary opponent learning techniques and create a configuration space of its strategic behavior.
To automatically configure this conceptually rich and highly parametric design, we use \gls{smac}, a general-purpose automated algorithm configuration procedure that has been used previously to optimize the performance of cutting-edge solvers for \gls{sat}, \gls{mip} and other NP-hard problems. 
We note that here, we apply automated algorithm configuration for the first time to a multi-agent problem.

The aim of this work is to automatically configure a negotiation algorithm with no fixed or pre-defined strategy. This agent can be configured to perform well on a user-defined set of training problem instances, with little restrictions on the size of the instances or instance sets. To demonstrate its performance, we configure the agent in an attempt to win an \gls{anac}-like bilateral tournament.

We show that we can win such a tournament with a comfortable margin of 5.1\% in increased negotiation payoff compared to the number two. These margins are not observed in a tournament without our negotiation agent, where the winning strategy obtains a marginal improvement in negotiation payoff of 0.012\%.

\section{Related work}\label{sec:related}
In this section, we discuss related work in the field of automated algorithm configuration, as well as some past applications in the research area of automated negotiation.

\subsection{Automated algorithm configuration}
In literature, automated algorithm configuration is also referred to as parameter tuning or hyperparameter optimization (in machine learning). It can be formally described as follows: given a parameterized algorithm \(A\), a set of problem instances \(I\) and a cost metric \(c\), find parameter settings of \(A\) that minimize \(c\) on \(I\)~\cite{Hutter2011SequentialOptimization}. The configuration problem occurs for example in solvers for \gls{mip} problems~\cite{Hutter2010AutomatedSolvers}, neural networks, classification pipelines, and every other algorithm that contains performance-relevant parameters.

These configuration problems can be solved by basic approaches such as manual search, random search, and grid search, but over the years researchers developed more intelligent methods to obtain the best possible configuration for an algorithm. Two separate part within these methods can be identified: how new configurations are selected for evaluation and how a set of configurations is compared.

F-Race~\cite{Birattari2010F-RaceOverview} races a set of configurations against each other on an incremental set of target instances and drops low performing configurations in the process. This saves computational budget, as not all configurations have to be tested on the full target instance set. The set of configurations to test can be selected either manually, as a grid search, or at random. Balaprakash et al.~\cite{Balaprakash2007ImprovementRefinement} extended upon F-Race by implementing it as a model-based search~\cite{Zlochin2004Model-basedSurvey}, which iteratively models and samples the configuration space in search of promising candidate configurations.

ParamILS~\cite{hutter2009paramils} does not use a model, but instead performs a local tree search operation to iteratively find better configurations. Like F-Race, ParamILS is capable of eliminating low performing configurations without evaluating them on the full set of instances.

Another popular method of algorithm configuring is GGA~\cite{Ansotegui2009AAlgorithms}, which makes use of genetic programming to find configurations that perform well. This method does not model the configuration space and has no method to eliminate low performing configurations early.

The final method we want to mention is \gls{smac}, which is an algorithm configuration method that uses a random forest model to predict promising configurations. It also includes an early elimination mechanism for promising configurations by comparing them with a dominant incumbent configuration on individual problem instances.

\subsection{Automated configuration in negotiation agents}
Earlier attempts for solving the automated configuration problem in automated negotiation mostly used basic approaches, such as random and grid search. The only advanced method used to configure negotiation strategies is the genetic algorithm.

Matos et al.~\cite{Matos1998} encoded a mix of baseline tactics as an chromosome and deployed a genetic algorithm to find the best mix. They assumed perfect knowledge of the opponents preferences and their strategy is only tested against itself on a single negotiation scenario. 
Eymann~\cite{Eymann2001} encoded a more complex strategy as a chromosome with 6 parameters, again only testing its performance against itself and using the same scenario. 
Dworman et al.~\cite{Dworman1996} implement the genetic algorithm in a coalition game with 3 players, with a strategy in the form of a hard coded if-then-else rule. The parameters of the rule are implemented as a chromosome. The strategy is tested against itself on a coalition game with varying coalition values.
Lau et al.~\cite{lau2006evolutionary} use a genetic algorithm to explore the outcome space during a negotiation session, but do not use it to change the strategy.

\section{Preliminaries}\label{sec:preliminaries}
Automated negotiation is performed by software agents called parties, negotiation agents or simply agents. Agents that represent opposing parties in negotiation are also referred to as opponents. We focus solely on negotiations between two parties, which is known as bilateral negotiation. The software platform that we use for agent construction and testing is \gls{genius}~\cite{Lin12}, which contains all the necessary components to setup a negotiation, allowing us to focus solely on agent construction.

In this paper, we use the \gls{saop} as negotiation protocol, which is the formalization of the \gls{aop} in \gls{genius}. Here, agents take turns and at each turn either make an (counter) offer, accept the current offer, or walk away. This continues until one of the parties agrees, or a deadline is reached, which is set to 60 seconds in this paper (normalized to \(t \in [0, 1]\)).

Besides a protocol we need a set of opponent agents \(A\) to negotiate against and a set of scenarios \(S\) to negotiate over. We call the combination of a single opponent \(a \in A\) and a single scenario \(s \in S\) a negotiation setting or negotiation instance \(\pi \in \Pi = A \times S\).

\subsection{Scenario}
The negotiations in this paper are performed over 
multi-issue scenarios. Past research has already described on how to define and use such scenarios in automated negotiation \cite{raiffa1982art,Marsa-Maestre2014FromHandbook,Baarslag2014WhatStop}. We adopt these standards in this paper and describe them briefly.

An issue is a sub-problem in the negotiation for which an agreement must be found. It can be either numerical or categorical. The set of possible solutions in an issue is denoted by \(I\) and the Cartesian product of all the issues in a scenario forms the total outcome space \(\Omega\). An outcome is denoted by \(\omega \in \Omega\).

Every party has his own preferences over the outcome space \(\Omega\) expressed through a utility function \(u(\omega)\), such that \(U : \Omega \rightarrow [0, 1]\), where a score of 1 is the maximum. We refer to our own utility function with \(u(\omega)\) and to the opponents utility function with \(u_o(\omega)\). The negotiations are performed under incomplete information, so the utility of the opponent is predicted, which is denoted by \(\hat{u}_o(\omega)\).

Each scenario has a Nash bargaining solution~\cite{Nash1950} that we will use for performance analyses. \autoref{eq:nash} defines this equilibrium.
\begin{equation}\label{eq:nash}
    \omega_{Nash} = \argmax_{\omega \in \Omega} \left(u(\omega) * u_{o}(\omega) \right)
\end{equation}

We simplify in this paper, by eliminating the reservation utility and discount factor from the scenarios for the experiments.

\subsection{Dynamic agent}\label{sec:DA}
We first create a Dynamic Agent with a flexible strategy equivalent to a configuration space. We implement a few popular components and add their design choices to the configuration space, increasing the chances that it contains a successful strategy. We refer to this configuration space (or strategy space) with \(\Theta\). We name the constructed agent Dynamic Agent \(DA(\theta)\), with strategy \(\theta \in \Theta\).

The dynamic agent is constructed on the basis of the BOA-architecture~\cite{Baarslag2014WhatStop}. We use this structure to give a brief overview of the workings of the dynamic agent and its configuration space.

\subsubsection{Bidding strategy}
The implemented bidding strategy applies a fitness value to the outcome space \(\Omega\) and selects the outcome with the highest fitness as the offer, which is an approach used by Lau et al.~\cite{lau2006evolutionary}. This fitness function \(f(\omega, t)\) balances between our utility, the opponent's utility and the remaining time towards the deadline. Such a tactic is also known as a time dependent tactic and generally concedes towards the opponent as time passes.

The fitness function in \autoref{eq:fitness} has three parameters:
\begin{itemize}
    \item A trade-off factor \(\delta\) that balances between the importance of our own utility and the importance of reaching an agreement.
    \item A factor to control an agents eagerness to concede relative to time, where \(e\). Boulware if \(0 < e < 1\), linear conceder if \(e = 1\), conceder if \(e > 1\).
    \item A categorical parameter \(n\) that sets the outcome where the fitness function concedes towards over time (\autoref{eq:fitness2}). Here, \(x^{last}\) is the last offer made by the opponent and \(x^+\) is the best offer the opponent made in terms of our utility.
\end{itemize}

\begin{equation}\label{eq:fitness}
\begin{split}
    f(\omega, t) & = F(t) * u(\omega) + (1 - F(t)) * f_n(\omega) \\
    F(t)         & = \delta  * (1 - t^\frac{1}{e})
\end{split}
\end{equation}
\begin{equation}\label{eq:fitness2}
\begin{split}
    f_1(\omega)  & = 1 - |\hat{u}_{o} (\omega) - \hat{u}_{o} (x^{last})| \\
    f_2(\omega)  & = \min(1 + \hat{u}_{o}(\omega) - \hat{u}_{o} (x^{last}), 1) \\
    f_3(\omega)  & = 1 - |\hat{u}_{o} (\omega) - \hat{u}_{o} (x^{+})| \\
    f_4(\omega)  & = \min(1 + \hat{u}_{o}(\omega) - \hat{u}_{o} (x^{+}), 1) \\
    f_5(\omega)  & = \hat{u}_{o}(\omega)
\end{split}
\end{equation}

\paragraph{Outcome space exploration}
The outcome space is potentially large. To reduce computational time and to ensure a fast response time of our agent, we apply a genetic algorithm to explore the outcome space in search of the best outcome. Standard procedures such as, elitism, mutation and uniform crossover are applied and the parameters of the genetic algorithm are added to the configuration space.

\paragraph{Configuration space}
The configuration space of the bidding strategy is summarized in \autoref{tab:hyperbidding}.

\begin{table}
    \centering
    \begin{tabular}{lll}
        \toprule
        \textbf{Description} & \textbf{Symbol} & \textbf{Domain} \\
        \midrule
        Trade-off factor & \(\delta\) & \([0, 1]\) \\
        Conceding factor & \(e\) & \((0, 2]\) \\
        Conceding goal & \(n\) & \(\{1,2,3,4,5\}\) \\
        Population size & \(N_{p}\) & \([50, 400]\) \\
        Tournament size & \(N_{t}\) & \([1, 10]\) \\
        Evolutions & \(E\) & \([1, 5]\) \\
        Crossover rate & \(R_c\) & \([0.1, 0.5]\) \\
        Mutation rate & \(R_m\) & \([0, 0.2]\) \\
        Elitism rate & \(R_e\) & \([0, 0.2]\) \\
        \bottomrule
    \end{tabular}
    \caption{Configuration space in bidding strategy}
    \label{tab:hyperbidding}
\end{table}

\subsubsection{Opponent model}\label{sec:OM}
The Smith Frequency model~\cite{VanGalenLast2012} is used to estimate the opponents utility function \(\hat{u}_o(\omega)\). According to an analysis by Baarslag et al.~\cite{Baarslag2013PredictingNegotiation}, the performance of this opponent modelling method is already quite close to that of the perfect model. No parameters are added to the configuration space of the Dynamic Agent.

\subsubsection{Acceptance strategy}
The acceptance strategy decides when to accept an offer from the opponent. Baarslag et al.~\cite{Baarslag2014a} performed an isolated and empirical research on popular acceptance conditions. They combined acceptance conditions and showed that a combined approach outperforms its parts. Baarslag et al.\ defined four parameters and performed a grid-search in search of the best strategy. We adopt the combined approach and add its parameters (\autoref{tab:hyperaccepting}) to the configuration space of the Dynamic Agent. For more details on the combined acceptance condition, see ~\cite{Baarslag2014a}.

\begin{table}
    \centering
    \begin{tabular}{lll}
        \toprule
        \textbf{Description} & \textbf{Symbol} & \textbf{Domain} \\
        \midrule
        Scale factor & \(\alpha\) & \([1, 1.1]\) \\
        Utility gap & \(\beta\) & \((0, 0.2]\) \\
        Accepting time & \(t_{acc}\) & \([0.9, 1]\) \\
        Lower boundary utility & \(\gamma\) & \(\{{MAX}^W, {AVG}^W\}\) \\
        \bottomrule
    \end{tabular}
    \caption{Configuration space in acceptance strategy}
    \label{tab:hyperaccepting}
\end{table}

\subsection{Problem definition}\label{sec:problemdefinition}
The negotiation agents in the \gls{genius} environment are mostly based on manually configured strategies by competitors in \gls{anac}. These agents almost always contain parameters that are set by trial and error, despite the abundance of automated algorithm configuration techniques (e.g. Genetic Algorithm~\cite{Holland1992AdaptationIntelligence}). Manual configuration is a difficult and tedious job due to the dimensionality of both the configuration and the negotiation problem space.

A few attempts were made to automate this process as discussed in \autoref{sec:related}, but only on very specific negotiation settings with few configuration parameters. The main reason for this, is that many automated configuration algorithms require to evaluate a challenging configuration on the full training set. To illustrate, evaluating the performance of a single configuration on the full training set that we use in this paper would take \~18.5 hours, regardless of the hardware due to the real-time deadline. These methods of algorithm configuration are therefore impractical.

\paragraph{Automated strategy configuration}
We have an agent called Dynamic Agent \(DA(\theta)\), with strategy \(\theta \). We want to configure this agent, such that it performs generally well, using automated configuration methods. More specifically, we want the agent to perform generally well in bilateral negotiations with a real time deadline of \(60 [s] \). To do so, we take a diverse and large set of both agents \(A_{train}\) of size \(|A_{train}| = 20\) and scenarios \(S_{train}\) of size \(|S_{train}| = 56\) that we use for training, making the total amount of training instances \(|\Pi_{train}| = |A_{train}| * |S_{train}| = 1120\). Running all negotiation settings in the training set would take \(1120\) minutes or \(\sim 18.5\) hours, regardless of the hardware as we use real time deadlines.

Now suppose we have a setting for the Dynamic Agent based on the literature \(\theta_{l}\) and a setting that is hand tuned based on intuition, modern literature and manual tuning \(\theta_{m}\) that we consider baselines. Can we automatically configure a strategy \(\theta_{opt} \in \Theta\) that outperforms the baselines and wins an \gls{anac}-like bilateral tournament on a never before seen test set of negotiation instances \(\Pi_{test}\)?

\section{Automated configuration}\label{sec:optimisation}
The goal of our work is to create an agent that can be configured to obtain a negotiation strategy that performs well in a given setting. This requires us to define what it mean for a strategy to perform well. 
An obvious performance measure is the utility \(o(\theta,\pi)\) obtained using strategy \(\theta \) in negotiation instance \(\pi \). As we are interested in optimizing performance on the full set of training instances rather than for a single instance, we define the performance of a configuration on an instance set as the average utility:
\begin{equation}\label{eq:performance}
    O(\theta, \Pi) = \frac{1}{|\Pi|} \cdot \sum_{\pi \in \Pi} o(\theta, \pi),
\end{equation}
\begin{conditions}
    o & utility of configuration \(\theta \) on instance \(\pi \) \\
    O & average utility of configuration \(\theta \) on instance set \(\Pi \) \\
    \theta \in \Theta & parameter configuration\\
    \pi& single negotiation instance consisting of opponent agent \(a \in A\) and scenario \(s \in S\), where \(\pi = \langle a, s \rangle \in \Pi \) \\
    \Pi& set of negotiation instances
\end{conditions}

As stated in \autoref{sec:problemdefinition}, automated configuration methods that require evaluation on the full training set of instances, thus requiring \autoref{eq:performance} to be calculated, are impractical for our application. A second component that influences the amount of required evaluations, is the mechanism that selects configurations for evaluation. This is not a straightforward problem, as the configuration space is large, and simple approaches, such as random search and grid search, suffer from the curse of dimensionality.

\subsection{SMAC}
To solve the problem defined in \autoref{sec:problemdefinition}, we bring \gls{smac}, a prominent, general-purpose algorithm configuration procedure~\cite{Hutter2011SequentialOptimization}, into the research area of automated negotiation. We note that \gls{smac} is well suited for tackling the configuration problem arising in the context of our study:
\begin{enumerate}
    \item It can handle different types of parameters, including real- and integer-valued as well as categorical parameters.
    \item It can configure on subsets of the training instance set, reducing the computational expense.
    \item It has a mechanism to terminate poorly performing configurations early, saving computation time. If it detects that a configuration is performing very poorly on a small set of instances (e.g., a very eager conceder), it stops evaluating and drops the configuration.
    \item It models the relationship between parameter settings, negotiation instance features and performance, which tends to significantly reduce the effort of finding good configurations.
    \item It permits straightforward parallelization of the configuration process by means of multiple independent runs, which leads to significant reductions in wall-clock time.
\end{enumerate}

\gls{smac} keeps a run history (\autoref{eq:runhistory}), consisting of a configuration \(\theta_i\) with its associated utility \(o_i\) on a negotiation instance that is modeled by a feature set \(\mathcal{F}(\pi)\). 
A random forest regression model is fitted to this run history, mapping the configuration space and negotiation instance space to a performance estimate \(\hat{o}\) (\autoref{eq:modelmap}). This model is then used to predict promising configurations, which are subsequently raced against the best configuration found so far, until an overall time budget is exhausted.
We refer the reader to~\cite{Hutter2011SequentialOptimization} for further details on \gls{smac}.

\begin{equation}\label{eq:runhistory}
    R = \{( \langle \theta_1,\mathcal{F}(\pi) \rangle ,o_1),\dots,( \langle \theta_n,\mathcal{F}(\pi) \rangle ,o_n)\}
\end{equation}
\begin{equation}\label{eq:modelmap}
    \mathcal{M} : (\Theta \times \Pi ) \rightarrow \hat{o}
\end{equation}

In order for \gls{smac} to be successful in predicting promising configurations, it requires an accurate feature description of the negotiation instances that captures differences in complexity between these instances.

\paragraph{Automated algorithm configuration}
Suppose we have a set of opponent agents \(A\) and a set of negotiation scenarios \(S\), such that combining a single agent \(a \in A\) and a single scenario \(s \in S\) creates a new negotiation setting or instance \(\pi \in \Pi \). Can we derive a set of features for both the opponent and the scenario that characterize the complexity of the negotiation instance?

We approach this question empirically, by analyzing if a candidate feature set helps the automated algorithm configuration method in finding better configurations within the same computational budget.

\section{Instance Features}\label{sec:features}
The negotiation instances consist of an opponent and a scenario. We will extract features for both component separately and then combine them as a feature set of an instance (\autoref{eq:featuremap}). This feature description is used to by the configuration method to predict promising strategies for our Dynamic Agent \(DA(\theta)\).

\begin{equation}\label{eq:featuremap}
    \mathcal{F} : \Pi \rightarrow (X_{sc} \times X_{opp})
\end{equation}

\subsection{Scenario features}\label{sec:scenariofeatures}
A negotiation scenario consists of a shared domain and individual preference profiles. Ilany et al.~\cite{Ilany2016AlgorithmNegotiation} specified a list of features to model a scenario that they used for strategy selection in bilateral negotiation. Although the usage differs in their paper, the goal to model the scenario is the same, so we will follow Ilany et al.. The features are fully independent of the opponents behavior. An overview of the scenario features is provided in \autoref{tab:scenariofeatures}.

\begin{table}
    \centering
    \resizebox{\columnwidth}{!}{
    \begin{tabular}{l l l l}
        \toprule
        \textbf{Feature type} & \textbf{Description} & \textbf{Equation} & \textbf{Notes} \\
        \midrule
        Domain & Number of issues & \(|I|\) &  \\
        Domain & \makecell{Average number of\\values per issue} & \(\frac{1}{|I|} \sum\limits_{i \in I} |V_i|\) &  \\
        Domain & \makecell{Number of possible\\outcomes} & \(|\Omega|\) &  \\
        Preference & \makecell{Standard deviation of\\issue weights} & \(\sqrt{\frac{1}{|I|} \sum\limits_{i \in I} {(w_i - \frac{1}{|I|})}^2}\) &  \\
        Preference & \makecell{Average utility of all\\possible outcomes} & \(\frac{1}{|\Omega|} \sum\limits_{\omega \in \Omega} u(\omega)\) & \makecell{denoted\\by \(u(\bar{\omega})\)} \\
        Preference & \makecell{Standard deviation utility\\of all possible outcomes} & \(\sqrt{\frac{1}{|\Omega|} \sum\limits_{\omega \in \Omega} {(u(\omega) - u(\bar{\omega}))}^2}\) &  \\
        \bottomrule
    \end{tabular}}
    \caption{Scenario features}\label{tab:scenariofeatures}
\end{table}

\subsection{Opponent features}\label{sec:opponentfeatures}
This section describes the opponent features in detail. For each opponent, we store both the \textit{mean} and the \textit{\gls{cov}} of all features.

\subsubsection{Normalized time}
The time \(t \in [0,1]\) it takes to reach an agreement with the opponent.

\subsubsection{Concession rate}
To measure how much an opponent is willing to concede towards our agent, we use the notion of \gls{cr} introduced by Baarslag et al.~\cite{Baarslag2011}. The \gls{cr} is a normalized ratio \(CR \in [0,1]\), where \(CR = 1\) means that the opponent fully conceded and \(CR = 0\) means that the opponent did not concede at all. By using a ratio instead of an absolute value (utility), the feature is disassociated from the scenario.

To calculate the \gls{cr}, Baarslag et al.~\cite{Baarslag2011} used two constants. The minimum utility an opponent has demanded during the negotiation session \(u_{o}(x_{o}^{-})\) and the \gls{fyu}, which is the utility that the opponent receives at our maximum outcome \(u_{o}(\omega^{+})\).

\begin{figure}
    \centering
    \includegraphics[]{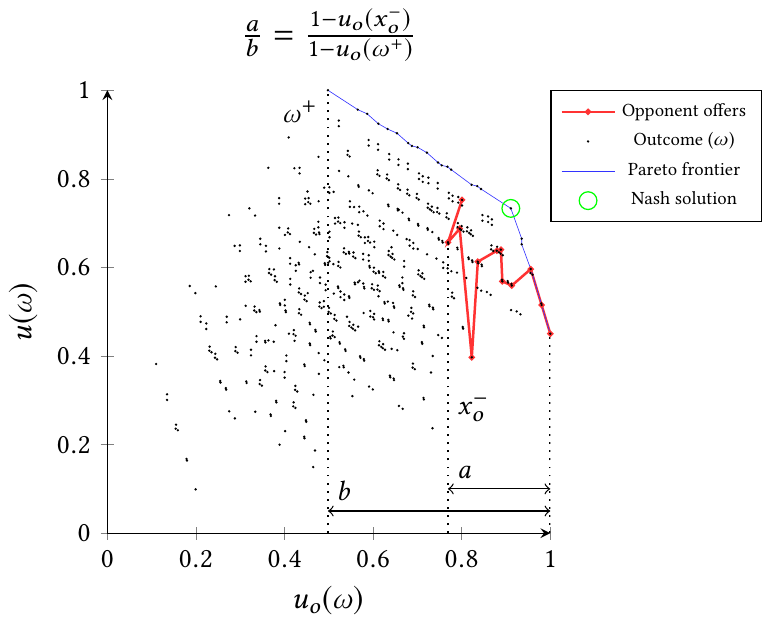}
    \caption{Visualization of \glsfirst{cr}}\label{fig:concessionrate}
\end{figure}

We present a formal description of the \gls{cr} in \autoref{eq:concessionrate} and a visualization in \autoref{fig:concessionrate}.
\begin{equation}\label{eq:concessionrate}
    CR(x_{o}^{-})= 
    \begin{cases}
        1 & \text{if } u_{o}(x_{o}^{-})\leq u_{o}(\omega^{+}), \\ 
        \frac{1 - u_{o}(x_{o}^{-})}{1 - u_{o}(\omega^{+})} & \text{otherwise.} 
    \end{cases}
\end{equation}

\subsubsection{Average rate}
\begin{figure}
    \centering
    \includegraphics[]{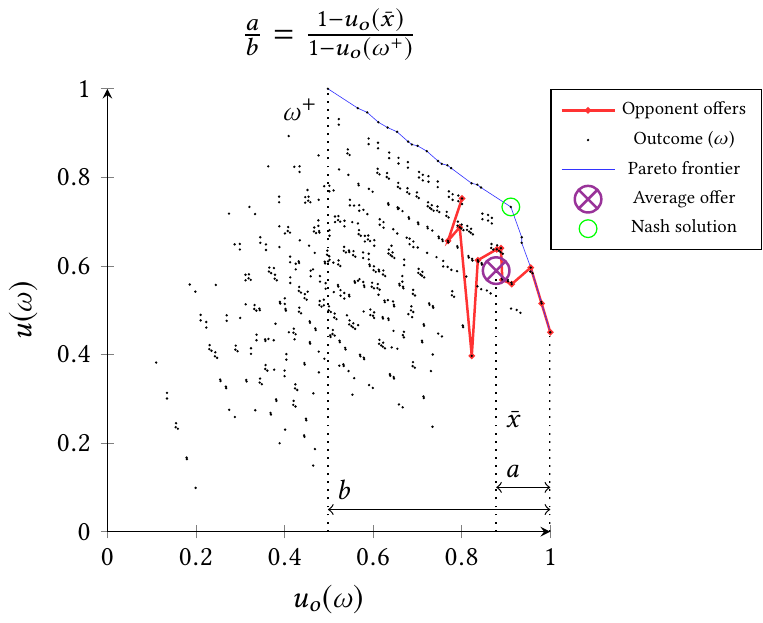}
    \caption{Visualization of \glsfirst{ar}}\label{fig:averagerate}
\end{figure}

We introduce the \gls{ar} that indicates the average utility an opponent has demanded as a ratio depending on the scenario. The two constants needed are the \gls{fyu} (\(u_o(\omega^{+})\)) as described in the previous section and the average utility an opponent demanded (\(u_o(\bar{x})\)).  The \gls{ar} is a normalized ratio \(AR \in [0,1]\), where \(AR = 0\) means that the opponent only offered his maximum outcome and \(AR = 1\) means that the average utility the opponent demanded is less than or equal to the \gls{fyu}. We present a definition of the \gls{ar} in \autoref{eq:averagerate} and a visualization in \autoref{fig:averagerate}.

\begin{equation}\label{eq:averagerate}
    AR(\bar{x})= 
    \begin{cases}
        1 & \text{if } u_o(\bar{x}) \leq u_{o}(\omega^{+}), \\ 
        \frac{1 - u_o(\bar{x})}{1 - u_{o}(\omega^{+})} & \text{otherwise.} 
    \end{cases}
\end{equation}

The \gls{ar} is another indication of competitiveness of the opponent based on average utility demanded instead of minimum demanded utility as the \gls{cr} is.

\subsubsection{Default configuration performance}
According to Hutter et al.~\cite{Hutter2011SequentialOptimization}, the performance of any default configuration on a problem works  well as a feature for that specific problem. For negotiation, this translates to the obtained utility of a hand-picked default strategy on a negotiation instance. The obtained utility is normalized and can be used as a feature for that negotiation instance. 

\begin{figure}
    \centering
    \includegraphics[]{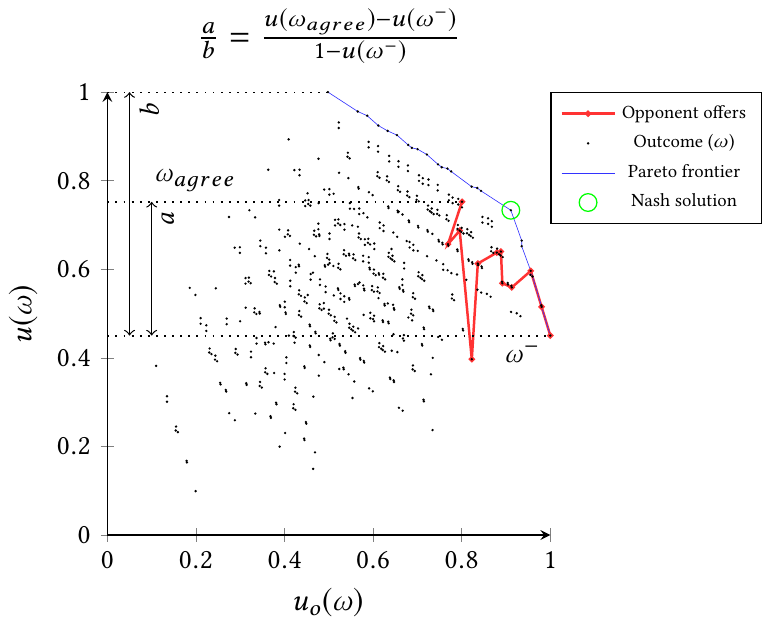}
    \caption{Visualization of \glsfirst{dcp}}\label{fig:defaultperf}
\end{figure}

We implement this concept as an opponent feature by selecting a default strategy and using it to obtain an agreement \(\omega_{agree}\) with the opponent. We then normalize the obtained utility and use it as the \gls{dcp} feature. We present the formal definition of this feature in \autoref{eq:defaultperf} and a visualization in \autoref{fig:defaultperf}.
\begin{equation}\label{eq:defaultperf}
    DCP(\omega_{agree})= 
    \begin{cases}
        0 & \text{if } u(\omega_{agree}) \leq u(\omega^{-}), \\ 
        \frac{u(\omega_{agree}) - u(\omega^{-})}{1 - u(\omega^{-})} & \text{otherwise.} 
    \end{cases}
\end{equation}

\subsection{Opponent utility function}
As can be seen in Figure \ref{fig:concessionrate}, \ref{fig:averagerate}, and \ref{fig:defaultperf}, the actual opponent utility function \(u_o(\omega)\) is used to calculate the opponent features. \gls{smac} is only used to configure the Dynamic Agent on the training set. As the opponent features are only used by \gls{smac}, we can safely use the opponent's utility function to construct those features (Equation \ref{eq:concessionrate}, \ref{eq:averagerate} and \ref{eq:defaultperf}) without giving the Dynamic Agent an unfair advantage during testing. The Dynamic Agent always uses the predicted opponent utility \(\hat{u}_(\omega)\) obtained through the model (\autoref{sec:OM}), as is conventional in the \gls{anac}. 

We provide an overview of when the predicted opponent utility function and when the actual opponent utility function is used in \autoref{tab:opputility}.

\begin{table}
    \centering
    \begin{tabular}{l | l l}
        \toprule
         & \textbf{Training} & \textbf{Testing} \\
        \midrule
        \(DA(\theta)\) & \(\hat{u}_o(\omega)\) & \(\hat{u}_o(\omega)\) \\
        \gls{smac} & \(u_o(\omega)\) & N/A \\
        \bottomrule
    \end{tabular}
    \caption{Opponent utility function usage}\label{tab:opputility}
\end{table}

\section{Empirical evaluation}\label{sec:experiment}
We must set baseline configurations to compare to the result of the optimization. The basis of our Dynamic Agent is derived from a paper by Lau et al.~\cite{lau2006evolutionary}. Though some functionality is added, it is possible to set our agent's strategy to resemble that of the original agent. We refer to this configuration from the literature as \(\theta_{l}\), its parameters can be found in \autoref{tab:baselineopt}.

Another baseline strategy is added, which is configured manually, as the literature configuration is outdated. A combination of intuition, past research, and manual search, is used for this manual configuration, which we consider default method for current \gls{anac} competitors. We present the manually configured parameters \(\theta_{m}\) in \autoref{tab:baselineopt} and an explanation below:
\begin{itemize}
    \item \textit{Accepting}: The acceptance condition parameters of \(\theta_l\) set a pure \(AC_{next}\) strategy with parameters \(\alpha = 1,\ \beta=0\). Baarslag et al.~\cite{Baarslag2014a} performed an empirical research on a variety of acceptance conditions and showed that there are better alternatives. We set the accepting parameters of our configuration to the best performing condition as found by Baarslag et al.~\cite{Baarslag2014a}.
    \item \textit{Fitness function}: Preliminary testing showed that the literature configuration concedes much faster than the average \gls{anac} agent, resulting in a poor performing strategy. We set a more competitive parameter configuration for the fitness function by manual search, to match the competitiveness of the \gls{anac} agents.
    \item \textit{Space exploration}: The domain used in the paper has a relatively small set of outcomes. We increased the population size, added an extra evolution to the genetic algorithm and made some minor adjustments to cope with larger outcome spaces.
\end{itemize}
\begin{table}
    \centering
    \resizebox{\columnwidth}{!}{
    \begin{tabular}{l l l l l l l l l l l l l l}
        \toprule
         & \multicolumn{4}{l}{Accepting} & \multicolumn{3}{l}{Fitness function} & \multicolumn{6}{l}{Space exploration} \\
        \cmidrule(lr){2-5}\cmidrule(lr){6-8}\cmidrule(lr){9-14}
        \(\theta \)       & \(\alpha \)                   & \(\beta \)                           & \(t_{acc}\)                           & \(\gamma \) & \(n\) & \(\delta \) & \(e\)    & \(N_{p}\) & \(N_{t}\) & \(E\) & \(R_c\) & \(R_m\)  & \(R_e\)  \\
        \midrule
        \(\theta_{l}\)  & \(1\)                         & \(0\)                                & \(1\)                                 & \(MAX^W\)   & \(1\)       & \(0.5\)     & \(0.5\)  & \(200\)     & \(3\)        & \(3\) & \(0.6\) & \(0.05\) & \(0.1\)  \\
        \(\theta_{m}\) & \(1\)                         & \(0\)                                & \(0.98\)                              & \(MAX^W\)   & \(4\)       & \(0.95\)    & \(0.05\) & \(300\)     & \(5\)        & \(4\) & \(0.6\) & \(0.05\) & \(0.05\) \\
        \bottomrule
    \end{tabular}}
    \caption{Baseline configurations parameters}\label{tab:baselineopt}
\end{table}

\subsection{Method}
\gls{smac} is run in \textit{embarrassingly parallel} mode on a computing cluster by starting a separate \gls{smac} process on chunks of allocated hardware. \gls{smac} selects a negotiation instance and a configuration to evaluate on that instance and calls the negotiation environment \gls{genius} through a wrapper function.

\paragraph{Input}
The training instances were created by selecting a diverse set of opponents and scenarios from the \gls{genius} environment. The scenarios have non-linear utility functions and vary in competitiveness and outcome space size (between 9 and 400\,000). The scenario features were calculated in advance as described in \autoref{sec:scenariofeatures}, and the configuration space is defined in \autoref{sec:DA}.

The opponent features, as defined in \autoref{sec:opponentfeatures}, can only be gathered by performing negotiations against the opponents. We gather these features in advance by negotiating 10 times in every instance with the manual strategy \(\theta_m\).

\paragraph{Hardware \& configuration budget}
We perform 300 independent parallel runs of \gls{smac} for 4 hours of wall-clock time each, on a computing cluster running \gls{slurm}. To ensure consistent results, all runs were performed on Intel\textsuperscript{\textregistered} Xeon\textsuperscript{\textregistered} CPU, allocating 1 CPU core, with 2 processing threads and 12 GB RAM to each run of \gls{smac}.

\paragraph{Output}
Every parallel \gls{smac} process outputs its best configuration \(\theta_{inc}\) after the time budget is exhausted. As there are 300 parallel processes, a decision must be made on which of the 300 configurations to use. To do so, the \gls{smac} random forest regression model conform \autoref{eq:modelmap} is rebuild and used to predict the performance of every \(\theta_{inc}\). The configuration with the best predicted performance is selected as best configuration \(\theta_{opt}\).

\subsection{Results}
The configuration process as described is run three times without instance features and three times with instance features, under identical conditions. There is now a total of 8 strategies: 2 baselines \([\theta_{l}, \theta_{m}]\), 3 optimized without features \([\theta_1,\theta_2,\theta_3]\), and 3 optimized with features \([\theta_4,\theta_5,\theta_6]\). An overview of the final configurations is presented in \autoref{tab:configurations}.
\begin{table}
    \centering
    \resizebox{\columnwidth}{!}{
    \begin{tabular}{l l l l l l l l l l l l l l}
        \toprule
         & \multicolumn{4}{l}{Accepting} & \multicolumn{3}{l}{Fitness function} & \multicolumn{6}{l}{Space exploration} \\
        \cmidrule(lr){2-5}\cmidrule(lr){6-8}\cmidrule(lr){9-14}
        \(\theta \) & \(\alpha \) & \(\beta \) & \(t_{acc}\) & \(\gamma \) & \(n\) & \(\delta \) & \(e\) & \(N_{p}\) & \(N_{t}\) & \(E\) & \(R_c\) & \(R_m\) & \(R_e\) \\
        \midrule
        \(\theta_{l}\) & \(1\) & \(0\) & \(1\) & \(MAX^W\) & \(1\) & \(0.5\) & \(0.5\) & \(200\) & \(3\) & \(3\) & \(0.6\) & \(0.05\) & \(0.1\) \\
        \(\theta_{m}\) & \(1\) & \(0\) & \(0.98\) & \(MAX^W\) & \(4\) & \(0.98\) & \(0.05\) & \(300\) & \(5\) & \(4\) & \(0.4\) & \(0.05\) & \(0.05\) \\
        \(\theta_{1}\) & \(1.001\) & \(0.048\) & \(0.901\) & \(AVG^W\) & \(3\) & \(0.879\) & \(0.00183\) & \(345\) & \(10\) & \(4\) & \(0.437\) & \(0.003\) & \(0.176\) \\
        \(\theta_{2}\) & \(1.041\) & \(0.001\) & \(0.904\) & \(AVG^W\) & \(4\) & \(0.913\) & \(0.00130\) & \(384\) & \(5\) & \(4\) & \(0.431\) & \(0.126\) & \(0.198\) \\
        \(\theta_{3}\) & \(1.009\) & \(0.026\) & \(0.910\) & \(MAX^W\) & \(1\) & \(0.977\) & \(0.00113\) & \(361\) & \(2\) & \(5\) & \(0.279\) & \(0.181\) & \(0.072\) \\
        \(\theta_{4}\) & \(1.032\) & \(0.022\) & \(0.931\) & \(AVG^W\) & \(3\) & \(0.914\) & \(0.00429\) & \(311\) & \(8\) & \(3\) & \(0.251\) & \(0.082\) & \(0.132\) \\
        \(\theta_{5}\) & \(1.015\) & \(0.017\) & \(0.925\) & \(AVG^W\) & \(5\) & \(0.961\) & \(0.00105\) & \(337\) & \(5\) & \(3\) & \(0.192\) & \(0.090\) & \(0.138\) \\
        \(\theta_{6}\) & \(1.027\) & \(0.022\) & \(0.943\) & \(AVG^W\) & \(3\) & \(0.985\) & \(0.00227\) & \(283\) & \(7\) & \(4\) & \(0.294\) & \(0.057\) & \(0.156\) \\
        \bottomrule
    \end{tabular}}
    \caption{Configurations overview}\label{tab:configurations}
\end{table}

The obtained configurations are now analyzed with an emphasis on the following three topics: 
\begin{enumerate}
    \item The influence of the instance features on the convergence of the configuration process.
    \item The performance of the obtained configurations on a never before seen set of instances.
    \item The performance of the best configuration in an \gls{anac}-like bilateral tournament.
\end{enumerate}

\subsubsection{Influence of instance features}
To study the influence of the instance features on the configuration process, we compare the strategies obtained by configuring with features and by configuring without features. Only the training set of instances is used for the performance comparison, as we are purely interested in the convergence towards a higher utility. 

Every configuration is run 10 times on the set of training instances \(\Pi_{train}\) and the average obtained utility is calculated by \autoref{eq:performance}. The results are presented in \autoref{tab:optconvergence}, including an improvement ratio over \(\theta_{m}\).
\begin{table}
    \centering
    \begin{tabular}{l l l l}
        \toprule
        \(\theta \)     & \(O(\theta,\Pi)\) & \(\frac{O(\theta,\Pi)-O(\theta_{m},\Pi)}{O(\theta_{m},\Pi)}\) & \textbf{Description}       \\
        \midrule
        \(\theta_{l}\)  & 0.533               & -0.307                                                                & Literature                 \\
        \(\theta_{m}\) & 0.769               & 0                                                                     & Manually configured                 \\
        \(\theta_1\)      & 0.785               & 0.020                                                                 & Configured without features \\
        \(\theta_2\)      & 0.770               & 0.000                                                                 & Configured without features \\
        \(\theta_3\)      & 0.792               & 0.029                                                                 & Configured without features \\
        \(\theta_4\)      & 0.800               & 0.040                                                                 & Configured with features    \\
        \(\theta_5\)      & 0.816               & 0.060                                                                 & Configured with features    \\
        \(\theta_6\)      & 0.803               & 0.044                                                                 & Configured with features    \\
        \bottomrule
    \end{tabular}
    \caption{Performance of configurations on \(\Pi = \Pi_{train}\)}\label{tab:optconvergence}
\end{table}

\gls{smac} is capable of improving the performance of the Dynamic Agent above our capabilities of manual configuration. We observe that configuration without instance features potentially leads to marginal improvements on the training set. Finally, we observe that the usage of instance features leads to less variation in final configuration parameters (\autoref{tab:configurations}) and to a significant improvement of obtained utility.

\subsubsection{Performance on test set}
Testing the configurations on a never before seen set of opponent agents and scenarios is needed to rule out potential overfitting. We selected a diverse set of scenarios and opponents for testing, such that \(|\Pi_{test}| = |A_{test}| * |S_{test}| = 16 * 28 = 448\).

Every configuration is once again run 10 times on the set of training instances \(\Pi_{test}\) and the average obtained utility is calculated by \autoref{eq:performance}. The results are presented in \autoref{tab:optperformance}, including an improvement ratio over \(\theta_{m}\).

\begin{table}
    \centering
    \begin{tabular}{l l l l}
        \toprule
        \(\theta \)     & \(O(\theta,\Pi)\) & \(\frac{O(\theta,\Pi)-O(\theta_{m},\Pi)}{O(\theta_{m},\Pi)}\) & \textbf{Description}       \\
        \midrule
        \(\theta_{l}\)  & 0.563               & -0.261                                              & Literature                 \\
        \(\theta_{m}\) & 0.763               & 0                                                   & Manually configured                 \\
        \(\theta_1\)      & 0.779               & 0.021                                               & Configured without features \\
        \(\theta_2\)      & 0.760               & -0.004                                              & Configured without features \\
        \(\theta_3\)      & 0.774               & 0.015                                               & Configured without features \\
        \(\theta_4\)      & 0.792               & 0.038                                               & Configured with features    \\
        \(\theta_5\)      & 0.795               & 0.042                                               & Configured with features    \\
        \(\theta_6\)      & 0.789               & 0.034                                               & Configured with features    \\
        \bottomrule
    \end{tabular}
    \caption{Performance of configurations on \(\Pi=\Pi_{test}\)}\label{tab:optperformance}
\end{table}

It is now clear that strategy configuration without instance features is undesirable as it potentially leads to a worse performing strategy. Configuration with instance feature on the other hand, still leads to a significant performance increase on a never before seen set of negotiation instances.

\subsubsection{ANAC tournament performance of best configuration}
The strategy configuration method is successful in finding improved configurations, but the results are only compared against the other configurations of our Dynamic Agent. No comparison is yet made with agents build by \gls{anac} competitors. We now compare the performance of the best configuration that we found to the \gls{anac} agents in the test set of opponents. 

We select \(\theta_5\) as the best strategy based on performance on the training set and enter the Dynamic Agent in an \gls{anac}-like bilateral tournament with a 60 second deadline. The Dynamic Agent is combined with the test set of opponents and scenarios. Every combination of 2 agents negotiated 10 times on every scenario, for a total amount of 38080 negotiation sessions. The averaged results are presented in \autoref{tab:anacresults}. We elaborate on the performance measures found in the table:

\begin{itemize}
    \item \textbf{Utility:} The utility of the agreement.
    \item \textbf{Opp. utility:} The opponent's utility of the agreement.
    \item \textbf{Social welfare:} The sum of utilities of the agreement.
    \item \textbf{Pareto distance:} Euclidean distance of the agreement to the nearest outcome on the Pareto frontier in terms of utility.
    \item \textbf{Nash distance:} Euclidean distance of the agreement to the Nash solution in terms of utility (\autoref{eq:nash}).
    \item \textbf{Agreement ratio:} The ratio of negotiation sessions that result in an agreement.
\end{itemize}

\begin{table}
    \centering
    \resizebox{\columnwidth}{!}{
    \begin{tabular}{lllllll}
        \toprule
        \thead{Agent} & \thead{Utility} & \thead{Opp.\\utility} & \thead{Social\\welfare} & \thead{Pareto\\distance} & \thead{Nash\\distance} & \thead{Agreement\\ratio} \\
        \midrule
        RandomCounterOfferParty & \underline{0.440} & \textbf{0.957} & 1.398 & 0.045 & 0.415 & 1.000 \\
        HardlinerParty & 0.496 & \underline{0.240} & \underline{0.735} & \underline{0.507} & \underline{0.754} & 0.496 \\
        AgentH & 0.518 & 0.801 & 1.319 & 0.118 & 0.408 & 0.904 \\
        ConcederParty & 0.577 & 0.848 & 1.425 & 0.047 & 0.358 & 0.964 \\
        LinearConcederParty & 0.600 & 0.831 & 1.431 & 0.046 & 0.350 & 0.964 \\
        PhoenixParty & 0.625 & 0.501 & 1.125 & 0.263 & 0.468 & 0.748 \\
        GeneKing & 0.637 & 0.760 & 1.396 & 0.061 & 0.383 & 0.993 \\
        Mamenchis & 0.651 & 0.725 & 1.377 & 0.087 & 0.360 & 0.927 \\
        BoulwareParty & 0.662 & 0.786 & 1.448 & 0.043 & 0.319 & 0.968 \\
        Caduceus & 0.677 & 0.486 & 1.163 & 0.241 & 0.453 & 0.784 \\
        Mosa & 0.699 & 0.640 & 1.339 & 0.113 & 0.385 & 0.902 \\
        ParsCat2 & 0.716 & 0.671 & 1.386 & 0.108 & \textbf{0.286} & 0.904 \\
        RandomDance & 0.737 & 0.716 & \textbf{1.453} & \textbf{0.024} & 0.344 & 0.998 \\
        ShahAgent & 0.744 & 0.512 & 1.256 & 0.188 & 0.389 & 0.821 \\
        AgentF & 0.751 & 0.605 & 1.356 & 0.100 & 0.367 & 0.918 \\
        SimpleAgent & 0.756 & 0.437 & 1.194 & 0.212 & 0.470 & 0.801 \\
        \(DA(\theta_5)\) & \textbf{0.795} & 0.566 & 1.361 & 0.087 & 0.407 & 0.922 \\
        \bottomrule
    \end{tabular}}
    \caption{Bilateral \gls{anac} tournament results using \(DA(\theta_5)\) (bold = best, underline = worst)}\label{tab:anacresults}
\end{table}

Using the Dynamic Agent with \(\theta_5\) results in a successful negotiation agent that is capable of winning a \gls{anac}-like bilateral tournament by outperforming all other agents (two-tailed t-test: \(p < 0.001\)). It managed to obtain a \(\frac{0.795-0.756}{0.756} * 100\% \approx 5.1\% \) higher utility than SimpleAgent, the number two in the ranking, while also outperformed it on every other performance measure.

Since the presence of our agent in the tournament also influences the performance of other agents, we also ran the full tournament without our Dynamic Agent as a sanity check. The top 5 performers of this tournament are presented in \autoref{tab:anacresults2}, along with their margins over the respective next lower-ranking agent in terms of utility.

\begin{table}
    \centering
    \begin{tabular}{llL}
        \toprule
        \thead{Agent} & \thead{Utility} & \multicolumn{1}{l}{\thead{Margin}} \\
        \midrule
        Mosa        & 0.715 & 3.01\% \\
        ShahAgent   & 0.736 & 2.43\% \\
        RandomDance & 0.754 & 0.65\% \\
        AgentF      & 0.759 & 0.01\% \\
        SimpleAgent & 0.759 \\
        \bottomrule
    \end{tabular}
    \caption{Bilateral \gls{anac} tournament without \(DA(\theta_5)\)}\label{tab:anacresults2}
\end{table}

\section{Conclusion}\label{sec:conclusion}
The two main contributions of this work are (1) the success of automated configuration of negotiation strategies using a general-purpose configuration procedure (here: \gls{smac}), and (2) an investigation of the importance of the features of negotiation settings.

\subsection{Configuration}
Two baseline strategies were selected for our comparison. The first configuration, \(\theta_{l}\), is based on publications from which we derived the agent~\cite{lau2006evolutionary,Baarslag2014a}. The second configuration, \(\theta_{m}\), is configured based on intuition, recent literature and manual search, which we considered the default approach for current \gls{anac} competitors. In \autoref{sec:experiment}, we automatically configured our dynamic Agent using \gls{smac}.

The configuration based on earlier work \(\theta_{l}\)~\cite{lau2006evolutionary} performed poorly compared to the manually configured configuration \(\theta_{m}\), and achieved 26.1\% lower utility on our test set. The best automatically configured strategy \(\theta_{5}\) outperformed both baseline configurations and achieved a 4.2\% increase in utility compared to \(\theta_{m}\). From this, we conclude that the automated configuration method is successful in outperforming manual configuration.

Our experiments show that the automated configuration method can produce a strategy that can win an \gls{anac}-like bilateral tournament by a margin of 5.1\% (\autoref{tab:anacresults}). 
This is particularly striking when considering that without our agent, the winner of the same tournament beats the next-based agent only by a margin of 0.01\%.

\subsection{Features}
We consider a set of features that characterizes the negotiation scenario as well as the opponent.
Our empirical results indicate that when using the negotiation instance features, \gls{smac} is able to find good configurations faster. 

Overall, using \gls{smac} in combination with instance features leads to less variation in the parameter settings between the final configurations obtained in multiple independent runs (\autoref{tab:configurations}, \autoref{tab:optconvergence}), as well as significant and consistent performance improvement. Furthermore, our results show that automated configuration without features does not always outperform manual configuration. Therefore, we conclude that the instance features presented in this paper are a necessary ingredient for the successful automated configuration of negotiation strategies.

\subsection{Future work}
For this initial step towards automated configuration of negotiation agents, the negotiation scenarios were simplified by removing the reservation utility and the discount factor. Now that we have demonstrated that our general approach can be successful, additional validation should be performed in more complex and different negotiation environments.

Over the years, it became clear that there is no single best negotiation strategy for all negotiation settings~\cite{Lin12}. In this work, we have presented a method to automatically configure an effective strategy for a specific set of negotiation settings. However, if this set becomes too diverse, we inherently end up in a situation where the automatically configured best strategy may not perform too well. Future work should exploit the strategy space of the dynamic agent by extracting multiple complementary strategies for specific settings, along with an on-line selection mechanism that determines the strategy to be used in a specific instance.

\newpage

\end{document}